\documentclass[aps,prx,twocolumn,amsmath,amssymb]{revtex4} 
\usepackage{graphicx} 
\usepackage{Jonasmacros} 
\usepackage{bm}
\usepackage{mathptmx}
\usepackage{epstopdf} 

\usepackage{color}

\usepackage[colorlinks,citecolor=blue,linkcolor=Fuchsia]{hyperref}
\usepackage[usenames,dvipsnames,svgnames]{xcolor}

\usepackage{graphicx}
\usepackage{Jonasmacros}
\usepackage{bm}
\usepackage{mathptmx}
\usepackage{txfonts}
\newcommand{\tobedeleted}[1]{\textcolor{green}{#1}}
\renewcommand{\tobedeleted}[1]{\relax}

\renewcommand{\section}[1]{{\par\it #1.---}\ignorespaces}

\begin{document} 
\date{\today}

\title{Voltage induced switching dynamics of a coupled spin pair in a molecular junction}

\author{T. Saygun}

\author{J. Bylin}

\author{H. Hammar}

\author{J. Fransson}
\email{Jonas.Fransson@physics.uu.se}
\affiliation{Department of Physics and Astronomy, Box 516, 75120, Uppsala University, Uppsala, Sweden}

\begin{abstract}
Molecular spintronics is made possible by the coupling between electronic configuration and magnetic polarization of the molecules. For control and application of the individual molecular states it is necessary to both read and write their spin states. Conventionally, this is achieved by means of external magnetic fields or ferromagnetic contacts, which may change the intentional spin state and may present additional challenges when downsizing devices. Here, we predict that coupling magnetic molecules together opens up possibilities for all electrical control of both the molecular spin states as well as the current flow through the system. Tuning between the regimes of ferromagnetic and anti-ferromagnetic exchange interaction, the current can be, at least, an order of magnitude enhanced or reduced. The effect is susceptible to the tunnel coupling and molecular level alignment which can be used to achieve current rectification.
\end{abstract}

\maketitle

Molecular spintronics is a field which aims to merge the flexibility of synthetic design of molecular compounds with novel functionalities offered by magnetic properties in conjunction with electronics circuits \cite{NatMat.7.179}. Magnetically active molecules have been used to demonstrate spin valve effect using external magnetic fields \cite{NatMat.10.502}, stochastic switching between high and low conductive states by transitions between spin singlet and triplet ground states \cite{J.Chem.Phys.130.105101,Appl.Surf.Sci.254.7985,Phys.Rev.B.74.245320,NatNano.8.575}, controlled transport properties via paramagnetic atoms \cite{NanoLett.14.5365}, as well as their potential for quantum based computation \cite{Science.284.133,Nature.410.789,PhysRevLett.94.207208,NatNano.4.173,NatMat.12.337,NatMat.8.194}. Arrays of magnetic molecules inserted between conducting leads, moreover, provide an important forum to investigate fundamental magnetic properties of finite one-dimensional Ising or Heisenberg chains \cite{NatMat.13.782,Science.335.196,NatPhys.8.497} as well as potential for electrical and thermal control of the magnetic state. 
Certain classes of molecules, e.g., metal-phthalocyanines (MPc) and metal-porhyrins (MP) present chemical stability with specific optical and electrical properties make them highly appreciated for technological applications including organic field effect transistors \cite{ApplPhysLett.69.3066,ApplPhysLett.86.262109}, light emitting devices \cite{ApplPhysLett.74.3209,ApplPhysLett.72.2138} and photovoltaic cells \cite{ApplPhysLett.86.082106}, and for fundamental studies \cite{JPhysChemC.114.12258,ApplPhysLett.79.4148,AdvMater.18.320,JACS.127.12210,PhysRevB.77.035133,PhysRevLett.100.117601,NanoLett.14.5365}.

While incorporation of magnetic elements in molecular compounds can have a significant effect on the overall molecular transport properties \cite{NanoLett.14.5365}, the main established route to spintronics manipulations entails external magnetic fields \cite{NatMat.10.502} or ferromagnetic electrodes \cite{Science.332.1062,NatPhys.9.801}, often exploiting spin transfer torques from spin-polarized scattering \cite{NonEqNanoPhys} or Coulomb interaction \cite{PhysRevLett.90.166602}.
Here, we propose a different route to molecular spintronics based on voltage induced control of magnetic interactions that allows for all electrical control of the transport properties.
Deriving from local exchange interactions between the localized spin moments and the electrons in paramagnetic molecules, an indirect effective spin-spin interaction is generated between the molecular spin moments through the electron tunneling between the molecules \cite{PhysRevLett.113.257201}.
The tunneling electrons are, in turn, affected by the resulting spin state such that the system is driven into either a high or a low conductive regime, where the low conductive regime emerges from a novel form of spin singlet blockade phenomenon. Hence, we theoretically demonstrate that the voltage controlled magnetic interactions can be used to tune between regimes of high and low conductance in paramagnetic molecular dimers, without using external magnetic fields or ferromagnetic leads.
The effective spin-spin interaction is controlled by the energies of the highest occupied molecular orbitals (HOMO) and/or the lowest unoccupied molecular orbitals (LUMO) in the individual molecules and the intermolecular tunneling rate $\calT_c$. It is therefore possible to switch between high and low conductive states of the coupled molecules through variation of the electrical environment of the molecular structure, e.g., gating or voltage bias. Finally, we show that molecular level misalignment between the individual molecules leads to an asymmetry between the ferromagnetic and anti-ferromagnetic regimes with respect to the voltage bias which, in turn, gives rise to a suppression of the conductance in one direction of the current flow.

An important aspect of our work is the demonstration of non-equilibrium properties that cannot be predicted solely from an equilibrium consideration. While tuning between the ferromagnetic and anti-ferromagnetic regimes can be achieved using a gate voltage, this ability does not imply that the same property can be accomplished under non-equilibrium conditions. Thus, a generic conclusion of our results is that non-equilibrium aspects have to be considered to fully comprehend and control the physical properties.

\begin{figure}[t]
\begin{center}
\includegraphics[width=\columnwidth]{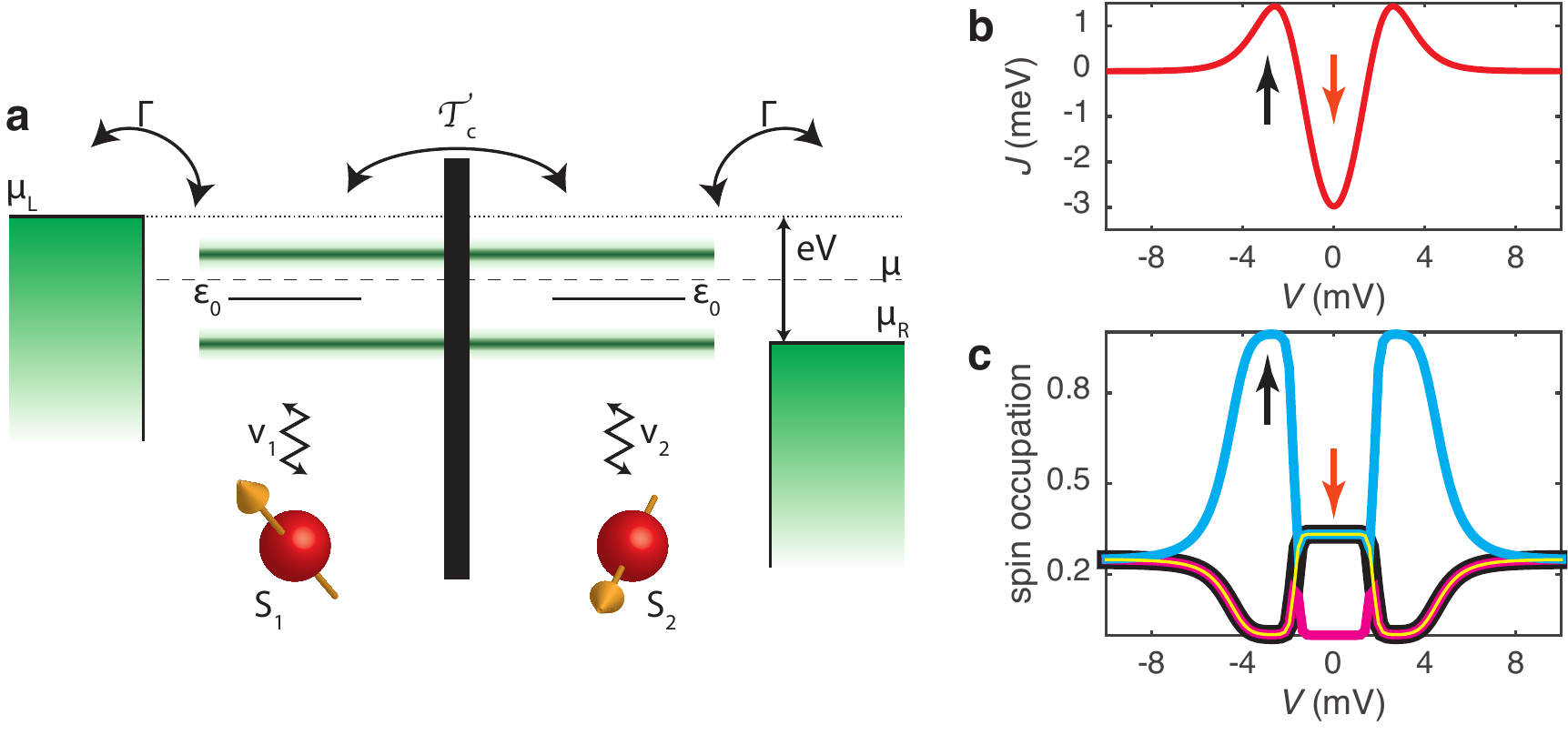}
\end{center}
\caption{
(a), Molecular dimer of paramagnetic molecules. An electron (at energy $\dote{0}$) in each molecule interacts with the localized spin moment ($\bfS_n$, $n=1,2$) via exchange ($v_n$), with the electron in the adjacent molecule (tunneling rate $\calT_c$), and with electrons in the left/right electrode (coupling $\Gamma$). The left/right non-magnetic electrode is characterized by its electrochemical potential ($\mu_{L/R}$). Effective molecular orbitals ($\dote{0}\pm\calT_c$) emerge from intermolecular tunneling.
(b), Effective exchange interaction between the localized spin moments as function of the voltage bias $V$.
(c), Occupation of the states of the spin dimer. The cyan curve represents the occupation of the lowest energy eigenstate of the spin  dimer which changes character between spin singlet and spin triplet states as function of the voltage bias. Other colors analogously represent the occupation of the consecutively higher energy eigenstates. Around equilibrium three states are degenerate and form the spin triplet.
Calculations have been made at $T=4$ K for $\dote{0}=0$, $\calT_c=1$ meV, $v_n=5$ meV, and $\Gamma=1$ meV.
In panels (b) and (c), the ferromagnetic and anti-ferromagnetic regimes of the spin dimer are indicated with red and black arrows, respectively.
}
\label{fig-Fig1}
\end{figure}


\section{Effective magnetic interactions}
Effective interactions between magnetic moments in solid materials as well as in chemical compounds where dipolar interactions can be ignored derive from local exchange interactions between the magnetic moment ($\bfS_m$) and the nearby electronic spin structure ($\bfs(\bfr)$) $\Hamil_K=\sum_m\int v(\bfr,\bfr_m)\bfs(\bfr)\cdot\bfS_md\bfr$, where $v(\bfr,\bfr_m)$ is the exchange integral between delocalized and localized spin $\bfs(\bfr)$ and $\bfS_m$, respectively. From this basic concept it can be deduced that the effective spin-spin interaction can be mapped onto the spin interaction Hamiltonian \cite{PhysRevLett.113.257201,PhysRevB.69.121303}
\begin{align}
\Hamil_S=&
	\sum_{mn}
	\biggl(
		J_{mn}\bfS_m\cdot\bfS_n
		+
		\bfS_m\cdot\mathbb{I}_{mn}\cdot\bfS_n
		+
		\bfD_{mn}\cdot[\bfS_m\times\bfS_n]
	\biggr)
	,
\label{eq-Hs}
\end{align}
where $J_{mn}$, $\bfD_{mn}$, and $\mathbb{I}_{mn}$ denote the Heisenberg, Dzyaloshinski-Moriya, and Ising interaction parameters, respectively. The interaction parameter depends on the properties of the electronic structure surrounding the localized spin moments.
Considering the interaction between different spins, $m\neq n$, it can be shown that the Heisenberg interaction, which is of isotropic nature, essentially depends on the charge density available to mediate the coupling between the spins and is generally finite in metallic materials as well as in the types of molecular compounds considered in a spintronics context.
In terms of Eq. (\ref{eq-Hs}), the spins tend to align ferromagnetically whenever $J_{mn}<0$ while an anti-ferromagnetic alignment is favored for $J_{mn}>0$.
The Ising interaction is anisotropic which is a property that stems from a spin-polarized electronic structure, without which it vanishes. Analogously to the Heisenberg interaction, negative (positive) Ising interaction leads to ferromagnetic (anti-ferromagnetic) spins along the spin quantization axis of the electronic structure.
Finally, the Dzyaloshinski-Moriya interaction, which is a source for spin non-collinearity, is finite only whenever both time-reversal and inversion symmetries are broken. Materials with finite spin-orbit coupling fulfill this requirement, however, in molecular compounds these two symmetries can be broken by using ferromagnetic electrodes under non-equilibrium conditions \cite{PhysRevLett.113.257201}.
Among the self-interactions, $m=n$, the Ising interaction provides an anisotropic dipolar \cite{NatPhys.9.801} field to the local spins provided that the surrounding electronic structure is spin-polarized. The Heisenberg interaction adds a constant shift of the total energy, since $\com{S_m^2}{\Hamil_S}=0$, while the contribution from the Dzyaloshinski-Moriya interaction vanishes since $\bfS_m\times\bfS_m=0$. Hence, both these two self-interactions are uninteresting for variations in the spin excitation spectrum.

\begin{figure*}[t]
\begin{center}
\includegraphics[width=0.7\textwidth]{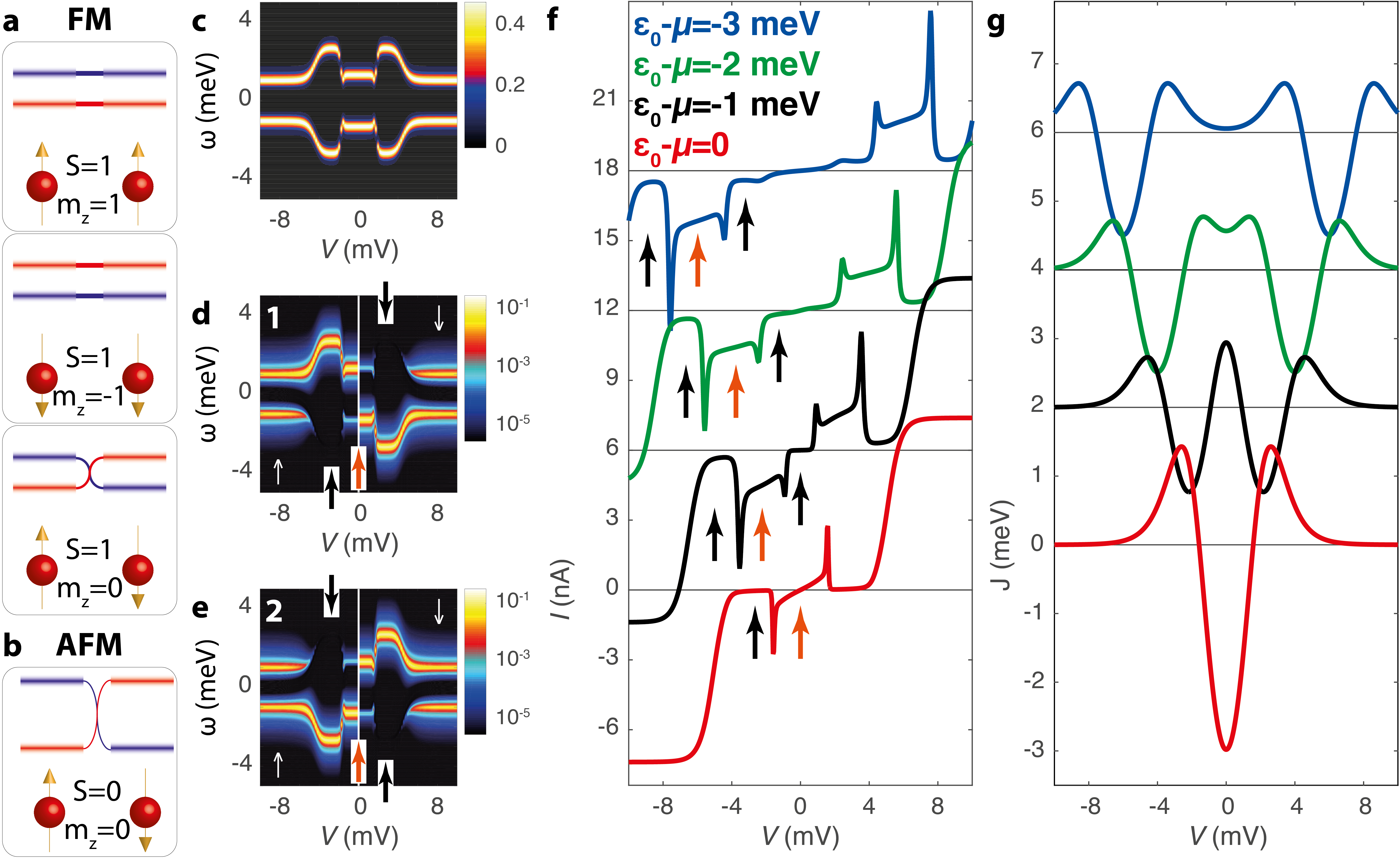}
\end{center}
\caption{
%
(a), The back-action of the spin triplet configurations on the molecular orbitals. The spin triplet configuration generates a delocalized DOS, resulting from the average of three degenerate spin configurations. 
(b), The back-action of the spin singlet configuration on the molecular orbitals, causes a strongly localized spin-projected DOS.
(c), Total density of electron states in the molecular dimer. 
(d) and (e), Molecule and spin-projected (indicated by white arrows) DOS of the left and right non-gated ($\mu=0$) molecules, respectively, as function of the voltage bias $V$ and energy $\omega$.
(f), Charge current through the molecular complex as function of the voltage bias ($V$), for different gating conditions ($\dote{0}-\mu=0,\ -1,\ -2,\ -3$ meV). 
(g), Corresponding effective exchange interactions. The plots are off-set for clarity (f), (g). 
%
In panels (d) -- (f), the ferromagnetic and anti-ferromagnetic regimes of the spin dimer are indicated with red and black arrows, respectively.
Parameters are as in Fig. \ref{fig-Fig1}.
}
\label{fig-Fig2}
\end{figure*}

Here, we consider a dimer of equivalent paramagnetic molecules inserted in the junction between metallic leads, see Fig. \ref{fig-Fig1} (a). We model this set-up using the Hamiltonian $\Hamil=\Hamil_M+\Hamil_\text{int}+\Hamil_L+\Hamil_R+\Hamil_T$. Here, $\Hamil_M=\sum_\sigma[\sum_{m=1,2}\dote{m}\ddagger{m\sigma}\dc{m\sigma}+\calT_c(\ddagger{1\sigma}\dc{2\sigma}+H.c.)]$, where $\ddagger{m\sigma}$ ($\dc{m\sigma}$) creates (annihilates) an electron in the $m$th molecule at the energy $\dote{m}=\dote{0}$ and spin $\sigma=\up,\down$, whereas $\calT_c$ defines the tunneling rate between the molecules. Internally in molecule $m$, the localized spin moment $\bfS_m$ interacts with the electron spin $\bfs_m=\sum_{\sigma\sigma'}\ddagger{m\sigma}\bfsigma_{\sigma\sigma'}\dc{m\sigma'}/2$ via exchange $\Hamil_\text{int}=\sum_mv_m\bfs_m\cdot\bfS_m$, where $v_m$ is the exchange integral, and we assume that $v_m=v$.
We focus on the case with non-magnetic leads, $\Hamil_{L/R}=\sum_{\bfk\sigma\in L/R}\dote{\bfk}\cdagger{\bfk}\cc{\bfk}$, where $\cdagger{\bfk}$ creates an electron in the left ($L$; $\bfk=\bfp$) or right ($R$; $\bfk=\bfq$) lead at the energy $\dote{\bfk}$ and spin $\sigma$.
Tunneling between the leads and molecules is described by $\Hamil_T=\sum_{\bfp\sigma}T_L\cdagger{\bfp}\dc{1\sigma}+\sum_{\bfq\sigma}T_R\cdagger{\bfq}\dc{2\sigma}+H.c.$.
Hence, $\Hamil_L+\Hamil_R+\Hamil_T+\Hamil_M$ provides a spin-degenerate electronic structure which mediates the exchange interactions, which implies that both the Dzyaloshinski-Moriya and Ising interactions vanish ($\bfD_{mn}=0$, $\mathbb{I}_{mn}=0$) and we retain the isotropic Heisenberg interaction only.
In this way we treat the spin dimer as an equilibrium system, i.e., conserved number of particles for which the occupations of the states is given by the Gibbs distribution, which is influenced by the tunneling current that flows through the molecular complex.
This set-up pertains to, e.g., MPc and MP where M denotes a magnetic transition metal atom, e.g., Cr, Mn, Fe, Co, Ni, Cu, and can be realized in, for example, mechanically controlled break-junctions \cite{NanoLett.14.5365} and scanning tunneling microscope \cite{NanoLett.13.4840,NanoLett.15.4024}. Having such systems in mind, also justifies that we neglect spin-orbit coupling in the molecular orbitals, since such coupling essentially pertains to the $d$-electrons constituting the paramagnetic moment, and also that we consider the molecular levels in a single particle form, relevant for $s$- and $p$-electrons. In such set-ups, the effective magnetic interaction parameter $J$ between the two spins can be calculated using the expression, see the Supporting Information,
\begin{align}
J=&
	-\frac{\calT_c^2}{8\pi}
	v^2
	\Gamma
	\int
		\Bigl(f_L(\omega)+f_R(\omega)\Bigr)
		(\omega-\dote{0})
\nonumber\\&\hspace{2cm}\times
		\frac{(\omega-\dote{0})^2-\calT_c^2-(\Gamma/8)^2}{|(\omega-\dote{0}+i\Gamma/8)^2-\calT_c^2|^4}
	d\omega,
\end{align}
where $\Gamma=\sum_{\chi=L,R}\Gamma^\chi$, with $\Gamma^\chi=2\pi\sum_{\bfk\sigma\in\chi}T_\chi^2\delta(\omega-\dote{\bfk})$,
is the coupling to the leads, and $f_{L/R}(\omega)$ is the Fermi function at the chemical potential $\mu_{L/R}$ of the left/right lead such that the voltage bias applied across the junction can be defined as $V=(\mu_L-\mu_R)/e$. The voltage bias dependence of $J$ is plotted in Fig. \ref{fig-Fig1} (b) for the case with $\dote{0}=0$ (for other parameters, see the figure caption). Near equilibrium ($V\approx0$), $\mu_R\approx\mu_L=\mu$, this integral gives a negative (ferromagnetic) value for $J$ whenever $\mu$ lies between the upper and lower effective orbital energies $\dote{0}\pm\calT_c$ of the molecular dimer, indicated by the red arrow in Fig. \ref{fig-Fig1} (b).
In this regime the ground state of the spin dimer is a spin triplet, labelled by $\ket{T;m=0,\pm}$ where $T$ denotes the triplet and $m$ its projection. This is illustrated in Fig. \ref{fig-Fig1} (c), which shows the occupation number for the four possible states in the spin dimer corresponding to the evolution of $J$ in panel (b). The red arrow indicates that the three projections of the spin triplet are equally occupied ($\sim1/3$).
With increasing voltage bias, $\mu_L$ and $\mu_R$ approach the orbital energies $\dote{0}\pm\calT_c$, which leads to a peaked positive (anti-ferromagnetic) $J$ as $\mu_{L/R}$ sweeps through the orbital energies (indicated by the black arrow in Fig. \ref{fig-Fig1} (b)). Hence, the spin dimer acquires a spin singlet ground state, $\ket{S}$, with nearly unit occupation as is indicated by the black arrow in Fig. \ref{fig-Fig1} (c).
By further increasing the voltage bias, eventually both energies $\dote{0}\pm\calT_c$ lie between $\mu_L$ and $\mu_R$ and $J$ remains positive but approaching zero, which leads to that the ground state of the spin dimer becomes a superposition of the spin triplet and spin singlet states. In this regime, where $J\rightarrow0$, the four states are equally occupied ($\sim1/4$) which can be seen in Fig. \ref{fig-Fig1} (c). It should be emphasized, however, that the total magnetic moment of the spin dimer vanishes for all voltages due to the absence of magnetic anisotropies.

An advantage with the present set-up, where we use non-spin polarized leads, compared to designs based on ferromagnetic leads is that the dipolar and quadrupolar fields considered in Ref. \citenum{NatPhys.9.801} here becomes vanishingly small. Therefore the effective electron mediated spin-spin interactions dominates the properties and control of the molecular dimer.

\section{Conductance states of molecular structure}
By tuning between the regimes with ferromagnetic and anti-ferromagnetic $J$, the state of the spin dimer is dynamically switched between spin triplet and singlet configurations. Due to the local interaction between the spin moment and the electrons, this switching directly affects the conductance through the molecular dimer. In fact, the state of the spin dimer influences the conducting orbitals and splits the molecular energy levels $\dote{0}\pm\calT_c$ into $E_\pm=\dote{0}\pm\widetilde{\calT}_c/2$, where $\widetilde{\calT}_c^2=v^2\av{S_1^z-S_2^z}^2+4\calT_c^2$ is the mean field splitting induced by the local spin moments, and $\av{S_m^z}$ is the molecule, or site, projected expectation value of the spin within the eigenbasis of the spin dimer. In the present case these expectation values take the values $\av{S_1^2}=-\av{S_2^z}=S$ and $S/3$ in the anti-ferromagnetic and ferromagnetic regimes, respectively, where $S$ denotes the spin moment. This anti-parallel alignment of the spins agrees with the very definition of the spin singlet state, as well as for the paramagnetic configuration $\ket{T,m=0}$ among the triplet states. The contributions to the expectation values from the ferromagnetic triplet states $\ket{T,m=\pm1}$ cancel each other, however.
Therefore, the ratio between the expectation values $|\av{S^z_1-S^z_2}|$ for the ferromagnetic and anti-ferromagnetic regimes is 1/3, which corresponds to the different distributions of the occupation numbers among the triplet and singlet states within the respective regime. Hence, the splitting of the molecular energy levels in the ferromagnetic regime is smaller than in the anti-ferromagnetic regime. This is illustrated in Fig. \ref{fig-Fig2} (c), where we plot the total density of electron states (DOS) in the two molecules. 
%
In comparison, conventional spintronics crucially depend on magnetic fields. In order to obtain the splitting between the conducting states $E_+$ and $E_-$, which here is induced by $v|\av{S_1^z-S_2^z}|$ with $v\sim0.5$ -- $20$ meV,\cite{ChemRev.104.5419,PhysRevLett.101.197208} one would have to apply a field strength of the order of $4|\av{S_1^z-S_2^z}|$ -- $170|\av{S_1^z-S_2^z}|$ T.

The main importance of the induced orbital splitting, however, is the distribution of the spectral weight of the molecular orbitals onto the individual molecules. This can be seen by analyzing the $2\times2$ matrix Green function (GF) for, e.g., molecule 1, $\bfG^{(1)}=\{G_{\sigma\sigma'}^{(1)}\}_{\sigma\sigma'=\up\down}$, which can be written
\begin{align}
\bfG^{(1)}(\omega)=&
	\frac{1}{2\widetilde\calT_c}
	\sum_{s=\pm}
		\frac{\widetilde\calT_c\sigma^0-s2v\av{S_2^z}\sigma^z}{\omega-E_s+i\Gamma/8},
\label{eq-GF1}
\end{align}
where $\sigma^z$ is a Pauli matrix whereas $\sigma^0$ is the identity.
In the limit of small $\calT_c$, the spectral weights of the component $G^{(1)}_{\up\up}$ around the energies $E_+$ and $E_-$ are $v\av{S_1^z-3S_2^z}-2\calT_c^2/v\av{S_1^z-S_2^z}$ and $v\av{S_1^z+S_2^z}-2\calT_c^2/v\av{S_1^z-S_2^z}$. Here, the former spectral weight is finite in both the anti-ferromagnetic and ferromagnetic regimes while the latter is finite only in the ferromagnetic regime and negligible in the anti-ferromagnetic. The distribution of the spectral weights for the component $G^{(1)}_{\down\down}$ is the opposite. Schematically, this is illustrated in Fig. \ref{fig-Fig2} (a),(b), while in Fig. \ref{fig-Fig2} (d) we plot the computed molecule and spin-projected DOS $-\im G^{(1)}_{\up\up(\down\down)}/\pi$ for positive (negative) voltage biases. Repeating the analysis for molecule 2, which GF is obtained from Eq. (\ref{eq-GF1}) by the replacements $\av{S_2^z}\rightarrow\av{S_1^z}$, $\sigma^z\rightarrow-\sigma^z$ and $\Gamma^L\rightarrow\Gamma^R$, shows that the distribution of the spectral weights is opposite to that of molecule 1, see Fig. \ref{fig-Fig2} (e), which shows the corresponding DOS for molecule 2. In particular this means that $G^{(1)}_{\sigma\sigma}$ and $G^{(2)}_{\sigma\sigma}$ have a negligible (finite) overlap, or in other words, the spin-projections of the electronic density are localized (delocalized), in the anti-ferromagnetic (ferromagnetic) regime, which has a large influence on the transport properties as we shall see next.

In the ferromagnetic regime, where the spin-projected DOS is delocalized in the molecular dimer, there are channels open for conduction which leads to a finite current flow. This is indicated by the red arrows in Fig. \ref{fig-Fig2} (f), where we plot the current $I$ for different gating conditions ($\dote{0}-\mu=0,-1,-2,-3$ mV), see the Supporting Information for details concerning the current. The corresponding exchange interactions $J$ are shown in Fig. \ref{fig-Fig2} (g). In the anti-ferromagnetic regime, on the other hand, the spin-projected DOS is localized and since we assume spin conservative tunneling between the molecules this leads to that an electron with spin $\sigma$ residing in molecule 1 has only a small probability to tunnel into molecule 2. Hence, the resulting current becomes severely suppressed, see Fig. \ref{fig-Fig2} (f) (black arrows).
The qualitative behavior of the current is the same in all four current traces, showing a high conductance in the ferromagnetic regime (red arrows) and a low conductance in the anti-ferromagnetic regime (black arrows) and we refer to the latter regime as a spin-singlet blockade. We stress the fact that since the dimer is constructed from paramagnetic molecules, this characteristics is independent of the spin quantization axis. The sharp current peaks separating the ferromagnetic and anti-ferromagnetic regimes result from a complete de-localization of the DOS when the exchange interaction parameter crosses between negative and positive values, which leads to a four fold degeneracy of the spin states.

With increasing voltage bias, the system evolves through regimes with different transport characteristics. First in the non-gated case ($\mu=0$), for small voltage biases the high conductance ferromagnetic regime (Fig. \ref{fig-Fig2} (g)) is, when increasing the voltage bias, subsequently followed by a low conductance state in the anti-ferromagnetic regime. As we saw above, the effective exchange interaction tends to become small by further increasing the voltage bias (Fig. \ref{fig-Fig2} (g)) such that the spin dimer evolves into a new regime with four-fold degenerate spin states. In this phase, the molecular orbitals are completely delocalized (Fig. \ref{fig-Fig2} (c), (d)) which permits an open flow of electrons through the molecular complex, yielding a significantly increased current, as can be seen in the current in the high bias regime (Fig. \ref{fig-Fig2} (f)).
Application of a gate voltage that shifts the energy levels with respect to $\mu$ leads to that the ferro- and anti-ferromagnetic regimes move to higher voltage biases (Fig. \ref{fig-Fig2} (f), (g)), since the molecular orbital energies may not lie on either side of $\mu$ in equilibrium but require a finite voltage bias to fulfill this condition. Therefore, for sufficiently large gating conditions, e.g., $|\dote{0}-\mu|\geq\calT_c$, which is fulfilled for $\dote{0}-\mu=-1$ meV, the system is anti-ferromagnetic in the low bias regime and only enter into the ferromagnetic regime for finite voltage biases (Fig. \ref{fig-Fig2} (f), (g)), which is followed by an another anti-ferromagnetic regime for a further increase of the voltage bias. This illustrates a generic property of the system, namely, that the ferromagnetic regime is surrounded by anti-ferromagnetic regimes which allows for switching between high and low conductance properties by shifting to either smaller or larger voltage biases -- a dual switching functionality. For even larger gating conditions, $\dote{0}-\mu\geq-2$ meV (Fig. \ref{fig-Fig2} (f), (g)), the system is initially in the highly conducting four-fold degenerate regime at low voltage biases and only thereafter evolves through the anti-ferromagnetic, ferromagnetic, and anti-ferromagnetic regimes, respectively, with increasing voltage bias. This behavior illustrates the systematic shift of the ferromagnetic and anti-ferromagnetic regimes away from equilibrium to higher voltage biases with gating.

\begin{figure}[t]
\begin{center}
\includegraphics[width=\columnwidth]{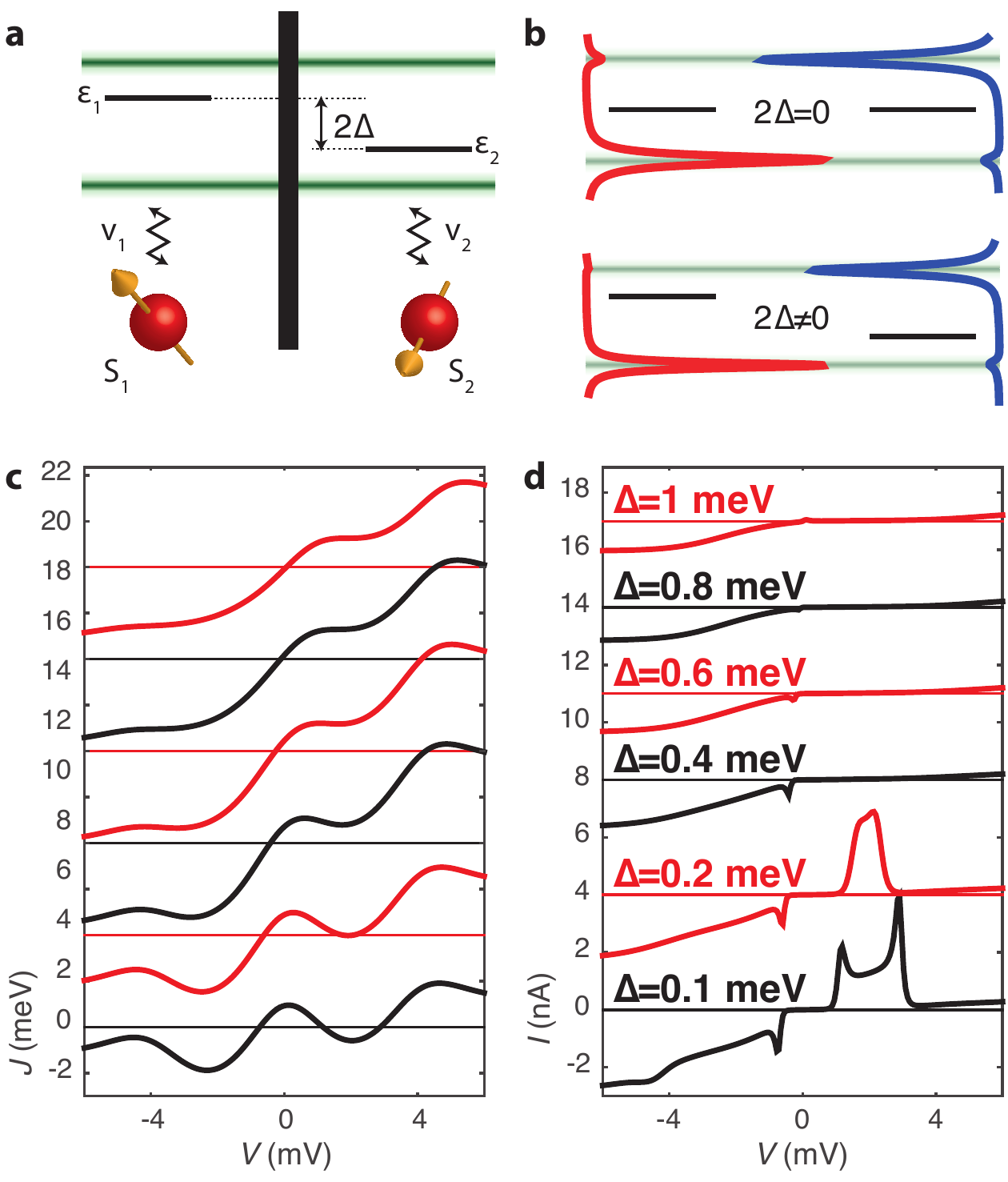}
\end{center}
\caption{
%
(a), Energy diagram of the non-equivalent molecules and the corresponding effective orbitals. The asymmetry is characterized by the level off-set $2\Delta$.
(b), Spatial distribution of the DOS for one spin-projection, illustrating the spatial asymmetry between the two molecules.
(c), Effective exchange interaction as function of the voltage bias for different asymmetric configurations of the molecular levels, $\dote{n}=\dote{0}+(-1)^n\Delta$, $n=1,2$, and $\Delta\in\{0.1,\ 0.2,\ 0.4,\ 0.6,\ 0.8,\ 1\}$ meV, for $\dote{0}-\mu=-1$ meV.
(d), Corresponding charge currents. 
Plots (c), (d) are off-set for clarity. Parameters are as in Fig. \ref{fig-Fig2}.
}
\label{fig-Fig3}
\end{figure}

\section{Non-equivalent molecules and rectification}
In the anti-ferromagnetic regime, the spin-projected DOS are strongly localized to either molecule (Fig. \ref{fig-Fig3} (b) -- upper panel), as discussed above.
The degree of localization can be further enhanced by replacing the molecular dimer with non-equivalent molecules in which the local electronic structure in each molecule is slightly different to one another (Fig. \ref{fig-Fig3} (a)). The imposed asymmetry can be used to amplify the singlet blockade and in this way to achieve current rectification. By introducing a finite level off-set between the orbital energies in the two molecules, the asymmetry of the spin-projected DOS can be fine tuned into almost complete localization (Fig. \ref{fig-Fig3} (b) -- lower panel).
The strongly asymmetric orbital configuration allows for fine tuning the effective exchange such that it becomes ferromagnetic for, e.g., negative voltage biases and anti-ferromagnetic for positive. This is illustrated in Fig. \ref{fig-Fig3} (c), which shows the effective exchange as function of the voltage bias for increasing level off-set $\Delta$ between the molecular orbital energies. By this separation of the ferromagnetic and anti-ferromagnetic regimes to negative and positive voltage biases, respectively, the system becomes an effective rectification device, which can be seen in the plots of the corresponding total currents $I$ shown in Fig. \ref{fig-Fig3} (d), also see the Supporting Information. Here, for small level off-sets $\Delta\leq0.2$ meV, the spin dimer has ferromagnetic regimes on both sides of zero voltage bias (Fig. \ref{fig-Fig3} (c)), such that the molecular system has a highly conducting range for negative and positive voltages. Moderately increasing the level off-set, $\Delta\gtrsim0.3$ meV completely removes the ferromagnetic regime for positive voltages which leads to an effective suppression of the current, hence, the system is strongly rectifying. We notice that rectification was recently robustly realized in molecular dimers \cite{NatNano.9.830}, however, arising from a completely different physical origin.

\section{Conclusions}
In conclusion we predict that electrical control of the effective exchange interaction in molecular spin dimer complexes can be utilized to provide switching function of the system by tuning the system between ferromagnetic and anti-ferromagnetic regimes, a tuning that can be provided by the voltage bias and gate voltage. In effect, the triplet and singlet spin states in the molecular dimer leads to a either delocalized or localized spin-projected DOS such that the ferromagnetic regime becomes highly conducting whereas the conductance is suppressed in the anti-ferromagnetic. This property opens up possibilities for electrical switching between different states associated with dramatic changes in the differential conductance. For molecular complexes with individual gating or non-equivalent paramagnetic molecules, it is predicted that the properties can be fine tuned for specific functional characteristics. Finite level off-set between the molecular orbitals can, for instance, be used to create effective current rectification where the high and low conductance phases are controlled by spin triplet and spin singlet states of the molecular spin dimer.
\\


The authors declare no competing financial interests.
\\

We thank A. V. Balatsky, O. Eriksson, A. Bergman for stimulating discussions and we acknowledge support from Vetenskapsr\aa det.

\newpage
\renewcommand{\figurename}{Figure}

\newcommand{\beginsupplement}{%
        \setcounter{table}{0}
        \renewcommand{\theequation}{S\arabic{equation}}%
        \setcounter{figure}{0}
        \renewcommand{\thefigure}{S\arabic{figure}.}%
     }
     
\beginsupplement
{\Large Supporting Information}
\\

\section{Model of the molecular spin dimer}
Consider two spin moments $\bfS_n$, $n=1,2$, embedded in a tunnel junction, each of which is coupled to a single level quantum dot via exchange. 
The molecular levels are coupled to one another through tunneling interactions and each of the molecules is coupled to an electron reservoir specified by their respective chemical potential $\mu_\chi$, $\chi=L,R$, where $L\ (R)$ denotes the left (right) reservoir. We can use a Hamiltonian of the following type to represent the interactions in the system:
\begin{align}
\Hamil=&
	\Hamil_L+\Hamil_R+\Hamil_T+\Hamil_M+\Hamil_{int}.
\end{align}
Here, $\Hamil_\chi=\sum_{\bfk\sigma}(\leade{\bfk}-\mu_\chi)\cdagger{\bfk}\cc{\bfk}$ represents the energy for the reservoir $\chi$ and we shall use $\bfp\ (\bfq)$ for the left (right) reservoir. The tunneling Hamiltonian $\Hamil_T=\Hamil_{TL}+\Hamil_{TR}$, where $\Hamil_{TL}=T_L\sum_{\bfp\sigma}\cdagger{\bfp}\dc{1\sigma}+H.c.$, and analogously for the right interaction. We will assume that the spin is conserved in the tunneling process. Further, the molecular dimer is represented by $\Hamil_M=\sum_{n=1,2,\sigma}\dote{n\sigma}\ddagger{n\sigma}\dc{n\sigma}+T_c\sum_\sigma(\ddagger{1\sigma}\dc{2\sigma}+H.c.)$, where the last terms denotes the tunneling coupling between the molecules. The interactions between the spins and electrons in the molecules are described by $\Hamil_\text{int}=\sum_nv_n\bfs_n\cdot\bfS_n$, where $\bfs_n=\sum_{\sigma\sigma'}\ddagger{n\sigma}\bfsigma_{\sigma\sigma'}\dc{n\sigma'}/2$. We assume that $v_n=v$.


\section{Exchange interaction}
Following the procedure described in Ref. \cite{PhysRevLett.113.257201}, we write the Heisenberg exchange interaction $J$ between the paramagnetic spin moments as
\begin{align}
J=&
	-\frac{\calT_c^2}{8\pi}
	v^2
	\sum_\sigma\Gamma_\sigma
	\int
		\frac{f_L(\omega)+f_R(\omega)}{|\omega-\dote{\sigma+}|^2|\omega-\dote{\sigma-}|^2}
\nonumber\\&\times
		(\omega-\dote{0})
		\frac{(\omega-\dote{0})^2-\calT_c^2-(\Gamma_{\bar\sigma}/4)^2}{|\omega-\dote{\bar\sigma+}|^2|\omega-\dote{\bar\sigma-}|^2}
	d\omega,
\label{eq-JH}
\end{align}
where the excitation energies $\dote{\sigma\pm}=(\dote{1\sigma}+\dote{2\sigma}\pm\Omega_\sigma-i\Gamma_\sigma/2)/2$, with $\Omega_\sigma^2=(\dote{1\sigma}-\dote{2\sigma}-i[\Gamma_\sigma^L-\Gamma_\sigma^R])^2+4\calT_c^2$, whereas $f_\chi(x)=f(x-\mu_\chi)$ is the Fermi function at the chemical potential of the electrode $\chi$. Here, also $\Gamma_\sigma=\sum_\chi\Gamma^\chi_\sigma$ where $\Gamma_\sigma^\chi=2\pi\sum_{\bfk\in\chi}T_\chi^2\delta(\omega-\dote{\bfk})$. In what follows we will assume that $\dote{n\sigma}=\dote{0}$ and $\Gamma^\chi_\sigma=\Gamma/4$. Then, $\Omega_\sigma^2=\calT_c$ and $\Gamma^\chi=\sum_\sigma\Gamma^\chi_\sigma=\Gamma/2$, $\Gamma_\sigma=\sum_\chi\Gamma^\chi_\sigma=\Gamma/2$, and $\Gamma=\sum_\chi\Gamma^\chi$. By reducing to the spin-degenerate set-up considered in the main text, Eq. (\ref{eq-JH}) simplifies to
\begin{align}
J=&
	-\frac{\calT_c^2}{8\pi}v^2\Gamma
	\int
		\Bigl(
			f_L(\omega)+f_R(\omega)
		\Bigr)
\nonumber\\&\times
		(\omega-\dote{0})
		\frac{(\omega-\dote{0})^2-\calT_c^2-(\Gamma/8)^2}{|(\omega-\dote{0}+i\Gamma/4)^2-\calT_c^2|^4}
	d\omega
\end{align}


\section{Tunneling current}
Considering the tunneling current flowing between the electrodes, we begin from the current in the left electrode, $I_L(t)=-e\dt\av{\sum_{\bfp\sigma}n_{\bfp\sigma}}$, which in the stationary regime leads to the expression
\begin{align}
I_L=&
	\frac{ie}{h}
	\sum_\sigma\Gamma^L_\sigma
		\int
			\biggl(
				f_L(-\omega)G_{\sigma\sigma}^{(1),<}(\omega)
				+
				f_L(\omega)G_{\sigma\sigma}^{(1),>}(\omega)
			\biggr)
		d\omega
		,
\end{align}
where $f_\chi(-\omega)=f(-\omega+\mu_\chi)=1-f_\chi(\omega)$.

The lesser and greater Green functions that appear in the expression for the tunneling current represents the electronic structure in molecule 1, where we take into account the coupling to the left lead and the paramagnetic spin $\bfS_1$. As we only consider the stationary regime, we treat the influence of the local spin moment as a mean field $\av{\bfS_1}=\av{S_1^z}$ acting on the molecular level $\dote{0}$. In absence of coupling between the spin channels, we can write the full retarded GF for the molecule 1 as
\begin{align}
\bfG^{(1)}(\omega)=&
	\frac{1}{2\widetilde\calT_c}
	\sum_{s=\pm}
		\frac{\widetilde\calT_c\sigma^0-s2v\av{S_2^z}\sigma^z}{\omega-E_s+i\Gamma/8},
\end{align}
where the excitation energies $E_\pm=\dote{0}\pm\widetilde\calT_c/2$ and $\widetilde\calT_c^2=v^2\av{S_1^z-S_2^z}^2+4\calT_c^2$.

The GF for the full system is a $4\times4$ matrix, $\mathbb{G}=\{\bfG^{(ij)}\}_{i,i=1,2}$, in which each entry is a $2\times2$ matrix $\bfG^{(ij)}=\{G^{(ij)}_{\sigma\sigma'}\}_{\sigma\sigma'}$. Here, the superscripts $i,j$ denotes the molecule 1 (2) if $ij=11$ (22) and the coupling between the molecules for $ij=12$ or $ij=21$. Each entry is defined by
\begin{align}
G_{\sigma\sigma'}^{(ij)}(\omega)=&
	\av{\inner{\dc{i\sigma}}{\ddagger{j\sigma'}}}(\omega)
\nonumber\\
	\equiv&
	\int_{-\infty}^t(-i)\av{\anticom{\dc{i\sigma}(t)}{\ddagger{j\sigma}(t')}}e^{i\omega(t-t')}dt'
	.
\end{align}
It turns out that the equation of motion for $\mathbb{G}$ can be written as the Dyson like equation
\begin{align}
\mathbb{G}=&
	\mathbb{G}_0+\mathbb{G}_0\bfSigma\mathbb{G},
\end{align}
where $\mathbb{G}_0$ is the bare GF for the coupled molecules but without coupling to the leads, whereas $\bfSigma$ defines this coupling. Now, since molecule 1 (2) only couples to the left (right) lead, the retarded form of this self-energy can be written as the diagonal matrix $\bfSigma=-i\, \diag{\Gamma^L_\up, \Gamma^L_\down, \Gamma^R_\up, \Gamma^R_\down}{}/2$. Thanks to the Dyson like equation for $\mathbb{G}$, the corresponding lesser and greater forms are given by the expression $\mathbb{G}^{</>}=\mathbb{G}^r\bfSigma^{</>}\mathbb{G}^a$, where the lesser and greater forms of the self-energy are given by
\begin{align}
\bfSigma^{</>}(\omega)=&
	(\pm i)
	\begin{pmatrix}
		f_L(\pm\omega)\begin{pmatrix} \Gamma^L_\up & 0 \\ 0 & \Gamma^L_\down \end{pmatrix} & \sigma^0 \\
		\sigma^0 & f_R(\pm\omega)\begin{pmatrix} \Gamma^R_\up & 0 \\ 0 & \Gamma^R_\down \end{pmatrix}
	\end{pmatrix}
	.
\end{align}

\begin{figure}[b]
\begin{center}
\includegraphics[width=0.99\columnwidth]{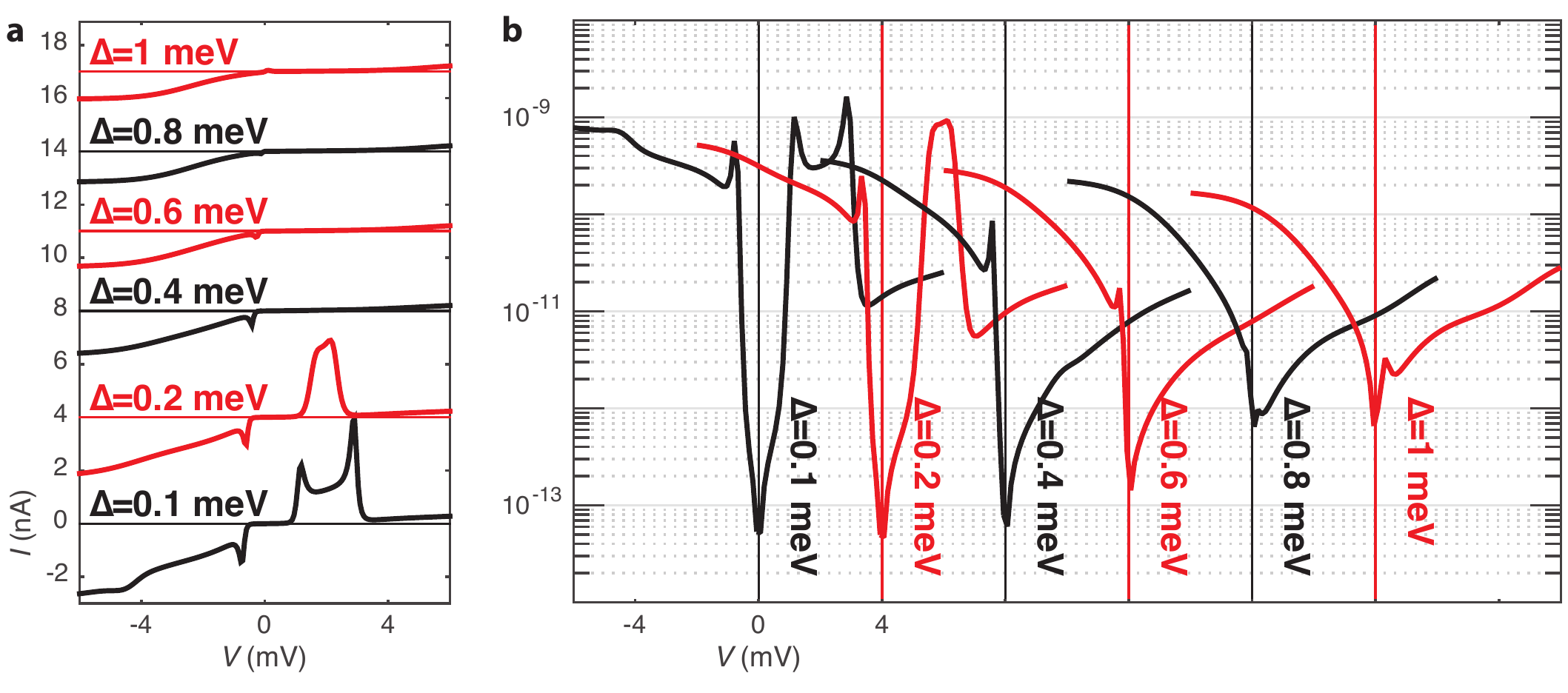}
\end{center}
\caption{Tunneling current for the asymmetric set-up discussed in the main text around Fig. 3. Panel (a) repeats the plots in Fig. 3 (d) of the main text while panel (b) provides the current on a logarithmic scale ($\log_{10}|I|$)}
\label{fig-Fig2SI}
\end{figure}

In Fig. \ref{fig-Fig2SI} we show the tunneling current for the asymmetric case discussed in the main text, where panel (b) displays the current amplitude on a logarithmic scale.

\section{External magnetic field}
Throughout this study we have relied on the absence of external magnetic fields, as such will introduce (Ising like) anisotropic components to the spin-spin exchange as well as undesired dipolar and quadupolar fields. Here, we can include the effects of the induced Ising like components which have a tendency to align the localized spin moments along the introduced anisotropy axis. Hence, under application of external magnetic field $\bfB=B_0\hat{\bf z}$ is can be noticed that while the low differential conductance is maintained in the anti-ferromagnetic regime, the differental conductance is actually increased in the ferromagnetic regime. This can be understood by considering that the spin singlet, which is occupied in the anti-ferromagnetic regime, is unaffected by the magnetic field. Hence, its influence on the electronic structure is essentially unaltered, i.e., the strong localization of the spectral weight is preserved. In the ferromagnetic regime, on the other hand, the Zeeman splitting of the spin triplet actually has a tendency to increase the delocalization of the spectral weight of the molecular electronic structure, something which then leads to an increased differential conductance.

It should be noticed, however, that a more thorough investigation of the expected effects under external magnetic fields have to include the dipolar and quadrupolar fields discussed in Ref. \cite{NatPhys.9.801}.

\section{Interactions}
In our model for the molecular electronic structure, $\Hamil_M$, we have neglected electron-electron interactions, as our model should pertain to the $s$- and $p$-orbitals in the structure. This is justified if one regards the molecular levels as being obtained from first principles calculations which often provide a reliable and sufficiently accurate single electron description of the HOMO and LUMO levels. One can, nevertheless, expect a finite repulsive interaction for these electrons which is not captured by the first principles calculations. In a mean field picture, there would likely not be any qualitative change of the molecular electronic structure as it essentially merely generates quantitative changes to the molecular spectrum.

\end{document}